\documentclass[english]{article}
\usepackage[latin1]{inputenc}
\usepackage{graphicx}
\usepackage{setspace}
\makeatletter

 \newcommand{\lyxaddress}[1]{
   \par {\raggedright #1 
   \vspace{1.4em}
   \noindent\par}
 }

\usepackage{babel}
\makeatother
\begin{document}

\title{On the limits of fractal surface behaviour in silica. A virtual adsorbates
simulation}

\author{Luis Guillermo Cota$^{\textrm{\dag}}$%
\thanks{Corresponding author. Electronic address: lgcota@gmail.com%
} , Pilar Alonso-Reyes$^{\textrm{\ddag}}$, Pablo de la Mora$^{\textrm{\dag}}$}

\maketitle

\lyxaddress{$^{\textrm{\dag}}$Departamento de Física. $^{\textrm{\ddag}}$Departamento
de Matemáticas. \\
Facultad de Ciencias. \\
Universidad Nacional Autónoma de México. \\
Circuito Exterior, Ciudad Universitaria. \\
México, D. F. 04510, México.}

\begin{abstract}
A computer simulation technique, suited to replicate real adsorption
experiments, was applied to pure simulated silica in order to gain
insight into the fractal regime of its surface. The previously reported
experimental fractal dimension was closely approached and the hitherto
uncharted lower limit of fractal surface behaviour is reported herein.
\end{abstract}

\section{Introduction}

Thanks to the pioneering adsorption experiments of Avnir \emph{et
al.,} it is now well established that a wide range of powdered substances
exhibit a rugged, or \emph{fractal} (as opposed to smooth, or Euclidean)
surface at the microscopic scale \cite{pfeifer83a,pfeifer83,avnir83,avnir84,pfeifer84,avnir85,rojanksy86,meyer86}. One distinctive principle behind fractality
is that the value of the measured property is dependent on the size
(and may also depend on the shape, \cite{vandamme86a}) of the measuring
instrument or ``yardstick'' (the historical issue, put forth in
the form of the celebrated question ``How long is the coast of
Britain?'', was originally raised, as we know it, by B. B. Mandelbrot
\cite{mandelbrot67}). Since in surface-area determination experiments
by adsorption, the yardsticks, or adsorbates, are usually small molecules,
the surface area reported is different for every adsorbate species
employed, and for all but the smoothest of surfaces. 

The unifying concept of \emph{fractal dimension} serves as a tool
to characterise intrinsic surface intricacies in a quantitative manner, 
and it necessarily implies that the
observed surface ruggedness remains invariant to changes of scale.
In terms of adsorption experiments, it means that any adsorbed molecule
will necessarily ``see'' a similar surface landscape, regardless
of its size and of the nature of the physisorption. In fact, topologic factors
seem to take precedence over the the former \cite{avnir84}.
 It is therefore possible to express the fractal dimension
of the substrate surface as the fractal dimension of a uniform coverage
of adsorbate molecules. The values reported range from close to 2
(purely Euclidean) for the smoothest surfaces, found, for example,
for  AEROSIL$^{\textregistered}$ ~silica spherules and graphite,
to close to 3, for the most convoluted (hard-to-imagine, almost volume-filling)
surfaces, found for some activated charcoals and silica gel \cite{avnir83,avnir84,pfeifer83a}.
(the authors in Ref. \cite{vandamme88}, on the other hand, present
an alternate view of what they call the ``ultimate fractal'',
the one with a fractal dimension of 3). 

The possibility of experimentally investigating with adsorption techniques
the fractal behaviour of a given material surface ultimately depends
on the size of the molecular probe employed. Therefore, adsorbates
ranging from nitrogen to coiled polymer molecules have been sucessfully
tested, covering a linear dimension range of about two orders of magnitude.
However, the actual limits of such fractal surface behaviour are difficult
to define. Since the lowest experimentally known bound is the effective
size of the nitrogen molecule, the purpose of this work is to explore
around this lower limit by resorting to simulated silica samples and
simulated ---''virtual''--- spherical adsorbates of adjustable, \emph{ad
hoc} radii.

\section{Experimental details}

\begin{figure}[h]
\includegraphics[%
  clip,
  scale=1.0]{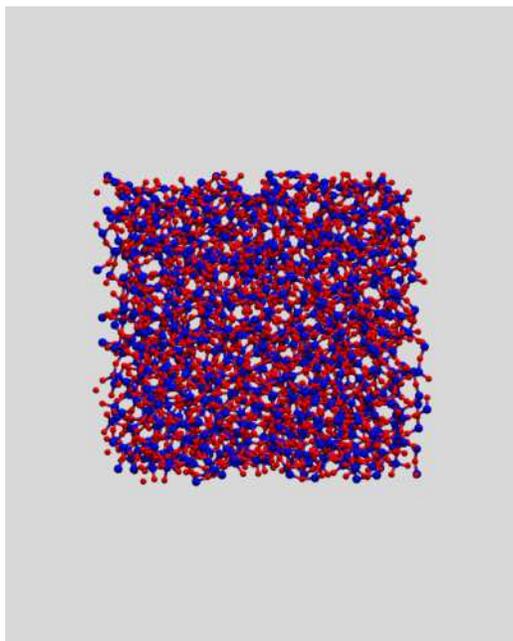}

\caption{\label{fig:silica3000}Representation of a 3000-atom vitreous SiO$_\textrm{2}$ cubic particle \cite{vink2003}.}
\end{figure}

The amorphous SiO$_\textrm{2}$ sample (``particle'') used consisted of a computational
box of cubic shape (165.5 $\textrm{Å}$ per side) containing 300,000 atoms,
  prepared by a modified Montecarlo method \cite{vink2003}. (Figure~\ref{fig:silica3000} shows a representation of a smaller silica system from
the same reference). This sample
reproduces very well the structure of a real system (with an overall
$T(r)$ discordance \cite{wright93,cao94} of 4.0\% between 1 and
10 $\textrm{Å}$).
It is pertinent to mention that, whereas the above simulated sample
is in excellent agreement with the bulk material, the surface of the
simulated particle was not specifically treated to achieve relaxation.
Therefore, the surface of our sample is equivalent to a virtual cut in 
the bulk without any perturbation on the coordination and energetics of 
the individual atoms. this is clearly only a first order approximation to 
the real surface. The preparation of a more ``physically correct'' surface 
large enough for the meaningful simulation of adsorption processes is a 
lengthy task not devoid of complications. This is in fact the matter of our
ongoing work.

\begin{figure}[h]
\includegraphics[%
  clip,
  scale=1.0]{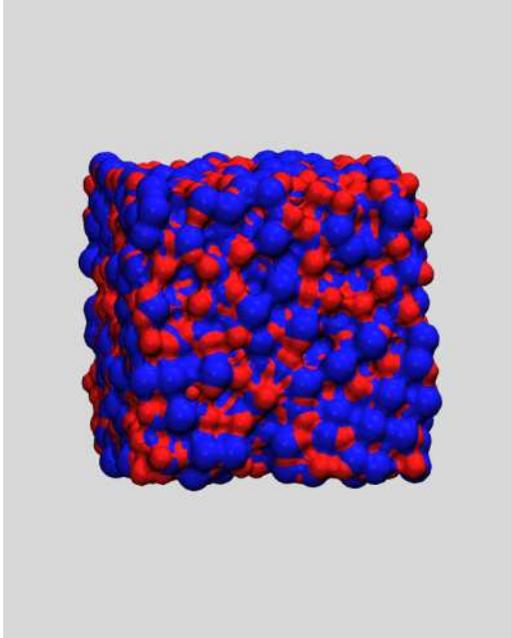}

\caption{\label{fig:superf}Arbitrary representation of the outer surface of the 
SiO$_\textrm{2}$ cubic particle from Figure~\ref{fig:silica3000}.} 
\end{figure}

\begin{figure}[h]
\includegraphics[%
  clip,
  scale=1.0]{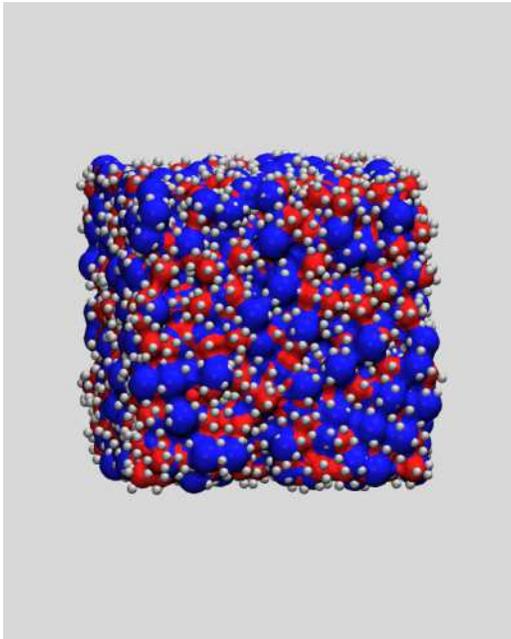}

\caption{\label{adsorc}Representation of the (as shown still incomplete) adsorption
process on the surface of the SiO$_\textrm{2}$ particle from 
Figure~\ref{fig:silica3000}.} 
\end{figure}

In real adsorption experiments, adsorbate molecules cover the surface
of a powdered substance usually in form of a monolayer.
 We have simulated
this process by randomly positioning spheres (the \textit{virtual
adsorbates)} of a given radius on the surface ---in fact, in any part
of the volume, as will be discussed below--- of the amorphous silica
sample (Figures~\ref{fig:superf} and \ref{adsorc}). In order for
a virtual adsorbate to be accounted as part of the monolayer two conditions
need to be met, namely that the adsorbate is not allowed to overlap
with other adsorbates and that it is not allowed to overlap with any
atom. (However, in order to save considerable calculation time, the
former ---quite severe--- restriction was forgone in favour of a maximum
overlap of 0.5\% between neighbouring adsorbates).

\begin{figure}[h]
\includegraphics[%
  clip,
  scale=0.5]{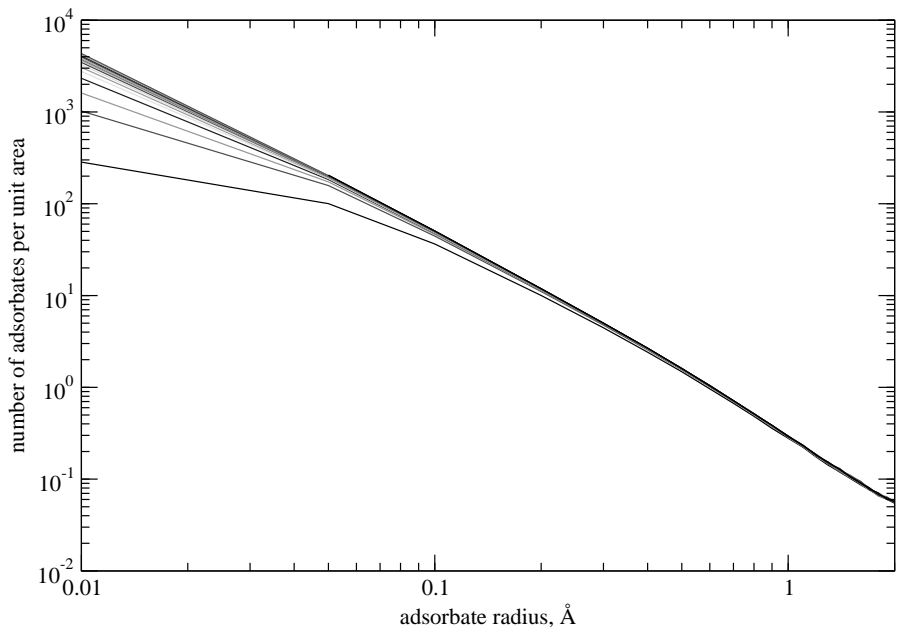}

\caption{\label{fig:sat_esp}Surface saturation curves for a 648 atom silica
conglomerate. The lower curve corresponds to 10$^{6}$ random
trials (equivalent to around 100 random trials per $\textrm{Å$^{\textrm{3}}$}$).
The upper curve corresponds to 10$^{\textrm{9}}$ random trials (10$^{\textrm{5}} $trials/$\textrm{Å$^{\textrm{3}}$}$).
Intermediate values are also shown. \emph{Unit area}, on the $y$ axis, refers to the surface of the computational box.}
\end{figure}

If a simulation is to reproduce the results of a real adsorption experiment,
it is important to assess the number of random trials large enough,
at any radius value within the explored range, to assure that the
silica particle under examination becomes fully covered, or \textit{saturated}
with virtual adsorbates. Failure to achieve surface saturation will
nonetheless yield a fractal dimension, although not that of the underlying
surface, since it is well known that the fractal dimension of a random
array of spheres depends on the occupation fraction of the array with
respect to the Euclidean space of the experiment \cite{hamburger96}.
Figure \ref{fig:sat_esp} shows the saturation behaviour of the simulated
adsorption process. Saturation is achieved when an increased number
of random trials is not able to increase the number of virtual adsorbates
on the surface. It is evident that even a rather small number of random
trials, 10$^{6}$, is able to produce a saturated surface at radius
values greater than around 0.3~$\textrm{Å}$ (for a small silica particle).
A much larger value, 10$^{10}$ (equivalent to 2200 random
trials per Å$^{3}$), was chosen for all production
runs.

\begin{figure}[h]
\includegraphics[%
  clip,
  scale=0.5]{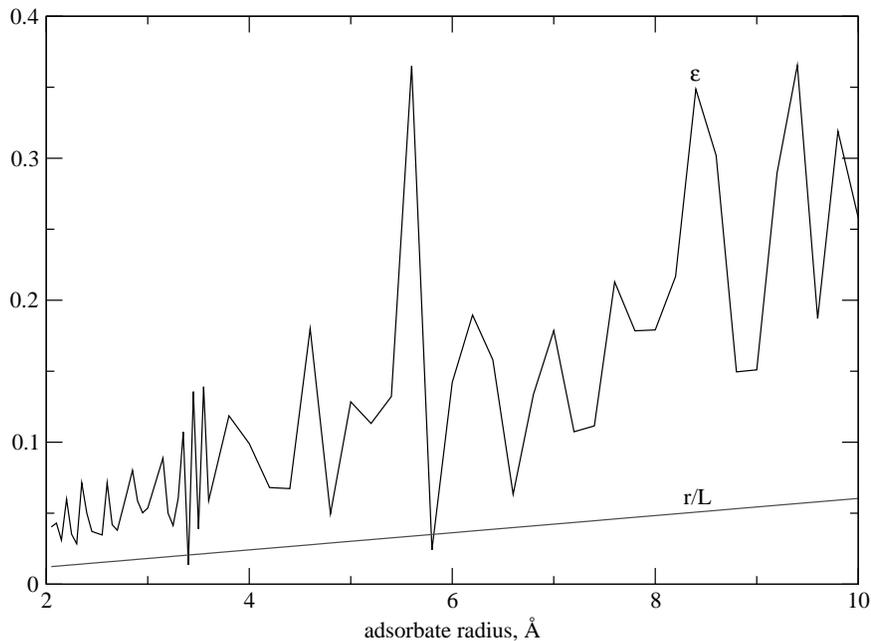}

\caption{\label{fig:error}Relative error, $\frac{n_{max}-n_{min}}{n_{avg}}$,
between four runs of the simulated adsorption process. Here, $r/L$
is the ratio between adsorbate radius and computational box (particle) 
edge length.}
\end{figure}

One of the strenghts of this method is that, once the particle surface
is saturated, the statistical error among several production runs
is rather small, at least in our range of interest. Figure \ref{fig:error}
shows this behaviour. However, statistical error increases with particle
radius, merely as a consequence of the $r/L$ ratio (the ratio between
adsorbate radius and particle edge length). In other words, adsorbates
become increasingly too big to accomodate on the particle surface.
This trend would eventually preclude a precise determination of the
upper cutoff of the fractal surface behaviour for this particle size
under the same conditions.

The fractal dimension of the surface is calculated once a series of
points representing the number of adsorbates versus adsorbate radius
is available, so that the trend can be accurately described,
within a definite range, by $n \propto r^{-d_f}$, where $n$ is the
number of adsorbates on the surface, $r$ is the adsorbate radius
and $d_f$ is the fractal dimension of the surface. 

It is pertinent to
note that this method relies on the fact that, as pointed out before, 
geometrical factors seem to predominate over the nature of the 
interaction energetics between the adsorbate and the substrate. Therefore,
no provision was made to account for the energetics involved.

In the simulation, the Shannon \& Prewitt's atomic radii (0.4 $\textrm{Å}$
for Si and 1.2 $\textrm{Å}$ for O) were used \cite{shannon65,shannon70}.

\section{Results and discussion}

\subsection{Fractal surface dimension }

\begin{figure}[h]
\includegraphics[%
  clip,
  scale=0.5]{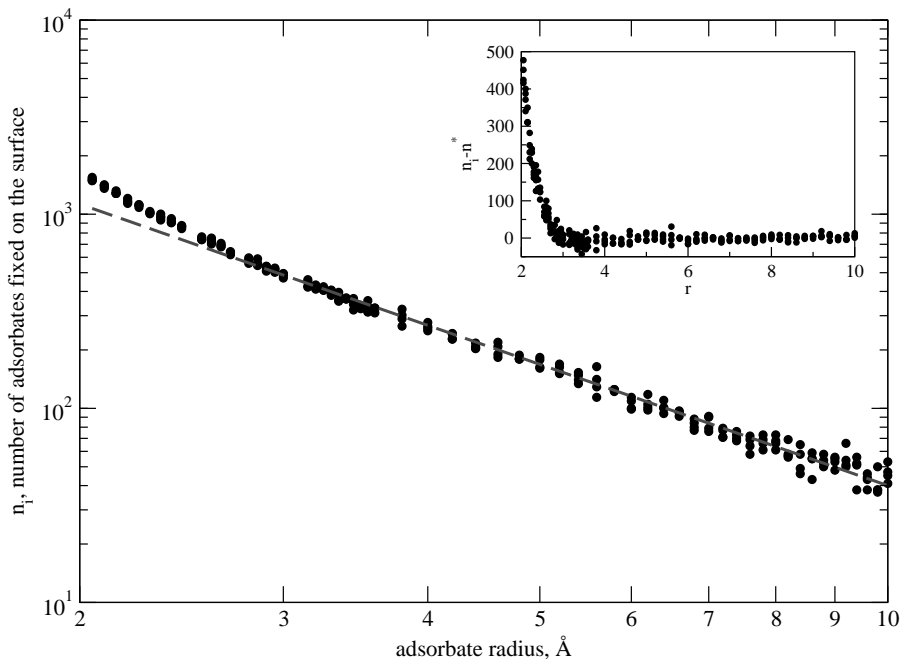}

\caption{\label{fig:df} \emph{Virtual adsorption curve,} log-log plot of
the $n(r)$~{\em vs.}~$r$ relationship for the virtual adsorption
experiment. A surface fractal behaviour is present wherein the linear
relationship holds. Here, the best fit is represented by the dashed
line. The fractal dimension, $d_f$, has herewith a value of 2.093
(after outlier exclusion, see text; the real experimental value is
$2.15 \pm 0.06$ \cite{avnir84}). \emph{Inset:} Departure from linearity 
---the lower cutoff value of surface fractal behaviour (see text)---, shown in linear scale. $N_i$
is the number of adsorbates attached to the surface during the computer
experiment at a given radius, $r_i$, and $n^*$ is the (best) fitted
value corresponding to such radius.}
\end{figure}

Figure \ref{fig:df}%
 shows the adsorption behaviour of our silica particle. A linear trend
is evident in most of the range studied (see inset). This is precisely
where a \emph{surface} fractal behaviour takes place, that can be
succintly described by the slope of this line, which is in fact the
\emph{fractal dimension} of the surface ($2 < d_f < 3$).

\begin{figure}
\includegraphics[%
  bb=-56bp 54bp 251bp 527bp,
  clip,
  scale=0.85]{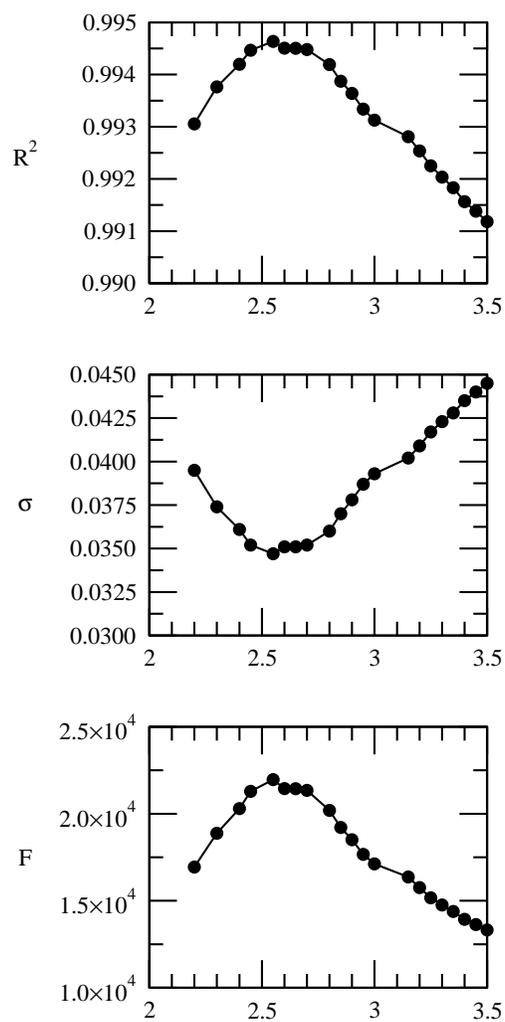}

\caption{\label{fig:ajustes} Location of the surface fractal behaviour
lower cutoff radius by use of the determination coefficient, $R^2$,
the standard deviation, $\sigma$, and the Fisher $F-test$ value,
respectively, from top to bottom. According to these figures of merit
the lower cutoff value is situated at 2.55$\textrm{Å}$.}
\end{figure}

Since it is not easy to assign merely by sight a lower cutoff value
of the linear trend (Inset, Figure~\ref{fig:df}), we resorted to
statistical multivariate analysis. Results are shown in Figure \ref{fig:ajustes}.

It is clear that every statistical measure of fit presented (minimum
standard deviation, maximum coefficient of determination and maximum
Fisher's $F-test$ value), is best at $r_c = 2.55$ $\textrm{Å}$.
This is, therefore, the lower cutoff of the fractal surface behaviour.
Our determination of the surface fractal dimension of amorphous silica
yields a value of $2.093 \pm 0.011$  ($R^2 = 0.996241$, $F = 29949$).
In order to stabilize the growing variance of the data (visible on
Fig.~\ref{fig:error}) we deemed suitable to run the analysis excluding
some of the most discrepant points, or \textit{outliers,} \textit{\emph{with
a standard rejection criteria of $\pm 2 \sigma$.}} In Ref. \cite{avnir84},
a value of $2.15 \pm 0.06$ is reported for ground Belgian quarz glass
of high purity, which is seemingly the closest physical match of our
simulated substance. The agreement is quite good, in spite of the
fact that our results refer to an \emph{unrelaxed} surface. Preliminary
surface fractal trial determinations of quartz make us suspect that
the fractal dimension is not very sensitive to the structural differences
between the surface of these polymorphs, and perhaps less so between
a relaxed and an unrelaxed surface. Studies are underway in order
to clarify this point.\\
On the other hand, the fractal dimension reported for crushed Corning lead
glass in Ref. \cite{avnir84},
$2.35 \pm 0.11$, rather than being an imprecise experimental determination
or a figure our present data should come close to, most likely suggests a 
yet-unexplored correlation between fractal dimension and glass composition. 
Perhaps a matter of future work. 

As regards the upper limit of the fractal surface behaviour, it is
evident that, if there is one, it seems not to be within the range
of our experiment. However, if the results from Ref. \cite{avnir84}
are recalled at this point, it is reasonable to overlap the ranges
covered by both sets of data: These authors have determined the fractality
of their silica in the range of yardstick cross-sectional areas of
16--10,600~$\textrm{Å}$$²$. However, since it is difficult to convert
molecular cross-section values to effective radius values \cite{meyer86},
a precise lower cutoff radius value cannot be readily extracted. Our data,
on the other hand, refer exclusively to spherical adsorbates, and
therefore, the cross-sectional area range can be directly obtained;
thus, we find that it spans roughly from 20--300~$\textrm{Å}$$²$ (corresponding to 2.55--10~Å).
Due to the good agreement between both sets of results, it can be inferred
that the fractal surface behaviour covers a yardstick cross-sectional
area range of \emph{at least} 20--10,600~$\textrm{Å}$$²$, which
implies that there is still surface fractality in silica at adsorbate radius
values of around 58~$\textrm{Å}$, considering a spherical experimental
molecule.

\subsection{Lower fractal limit and percolation onset}

Avnir and collegues \cite{hamburger96,lidar96,lidar97,malcai97} have
collected profuse theoretical and experimental evidence of the finite
range of fractal behaviour in many systems. On these grounds it is
well established that many ---if not most--- experimental systems present
a intrinsical ---that is, not experimentally bound--- fractal range
of around two decades in span. Fractal surface behaviour of silica,
is, most likely, no exception to this rule. However, to our knowledge,
existing experimental evidence is not enough to substantiate this
claim. 

We will next concentrate on the lower cutoff radius in order to correlate
it with a specific material characteristic of silica, namely its percolation
radius.

\begin{figure}[h]
\includegraphics[%
  clip,
  scale=0.5]{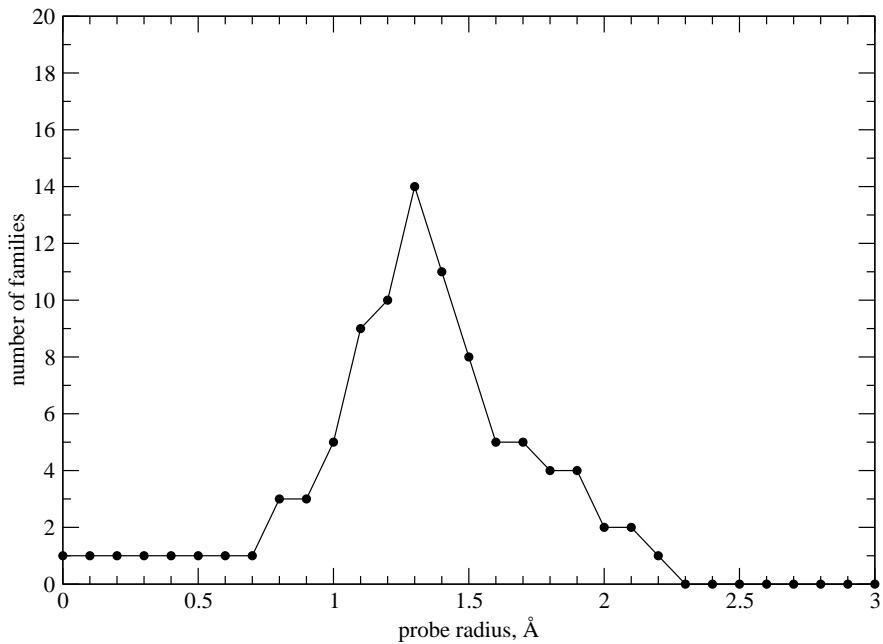}

\caption{\label{fig:numfam}Number of probe families in terms of probe radius.
\textit{Probe} is a spherical molecule inserted in an interatomic
cavity of the solid. \textit{Family} is an overlap of at least two
individual probes \cite{cota98}. Percolation is achieved when all
probes are connected, forming one big family. This graph is better
understood reading it from right to left.\protect \\
}
\end{figure}

Figure \ref{fig:numfam}, originally from Ref. \cite{cota98}, shows
the percolation behaviour of a simulated silica sample in terms of
the radius of ``probes'' inserted in cavities of the proper size
in the bulk of the material. In terms of physical experiments these
probes would be Argon atoms or nitrogen molecules, for example. %

It is pertinent to mention that the probe location algorithm from
Ref. \cite{cota98} is essentially similar to the one presented here:
while the former looks for overlapping (rather, connecting \cite{cota98})
probes within the structure, and the latter counts adsorbates on the
surface, both algorithms have a common core that looks for places
where to locate a virtual atom \textit{in the bulk.} In the latter
case, however, virtual adsorbates attach to outer atoms merely because
of size constraints and the absence of periodic boundary conditions.
As adsorbate size decreases, adsorbates start penetrating in the bulk,
and this is readily reflected in the slope of the \textit{n}~vs.~\textit{r}
curve (Fig. \ref{fig:df}). If this figure is now compared to Fig.
\ref{fig:numfam} above, it is evident that the changes observed in
both cases correspond to the onset of percolation, at around 2.55
$\textrm{Å}$ (keeping in mind the limitations of the former experiment),
which seems to become complete close to the value of 0.7~$\textrm{Å}$. 

Owing to the fact that the computing time required to continue the
experiment to even lower radii increases very abruptly with decreasing
radii, the adsorption experiment was finished at 2 $\textrm{Å}$.
As was outlined above, this value falls within the percolation transition
and, furthermore, the slope of the virtual adsorption curve at this
point is close to 3. However, it is to be expected that, at even lower
radii, another linear regime will be ultimately met, but now with
$d_f = 2$, corresponding to the surface of the spheres representing
the atoms themselves (if not flat, indeed smooth, therefore the
Euclidean value will be reached). It has been shown, on the other
hand \cite{hamburger96}, that $d_f$ is expected to take on the value
of the underlying Euclidean space of the experiment. Therefore, a
transition from $d_f$, corresponding to the fractal surface, to 3,
where the adsorbates fill up the internal cavity volume, and ultimately to 2,
 would probably be encountered with decreasing probe radii.

\section{Conclusions }

In order to gain some insight into the fractal regime of the pure
vitreous silica surface, we implemented a technique, a computational
analogue of real powder adsorption experiments. Our experiments produced
sucessful results compared with their real counterparts. 

A fractal dimension of $2.093 \pm 0.011$  was found, close to the
experimental value but with lower data dispersion, the latter being a desirable
and often-found characteristic in computer experiments. With the help
of statistical multivariate analysis, the existence of a lower surface fractal
behaviour cutoff was ascertained, and its precise value, 2.55 $\textrm{Å}$, was determined. It was
therefore possible to correlate this figure with the onset of percolation
behaviour in this material. The excellent agreement of our determination
of the fractal dimension of an unrelaxed simulated surface
with the fractal dimension of the real material might relate to the
insensitivity of the fractal dimension to surface relaxation.

\section{Acknowledgements}

Thanks are due to Jose J. Fripiat for helpful discussions at the early
stages of this work, to Alastair N. Cormack, to Xianlong Yuan and to Beyongwon
Park for help with the problematic of the neutron-diffraction code,
to Juan E. Murrieta for parallelising the virtual adsorbates code
and lastly to Ondrej Gedeon and to an anonymous referee for valuable 
critical remarks. The simulated
silica samples were kindly provided by Richard Vink. The generous
hosting of the DGSCA-UNAM to L. G. C. is hereby acknowledged. Partial
funding for this project was givent by UNAM's PAPIIT program.

The atomic representations were produced with the \emph{VMD} molecular
visualization suite \cite{HUMP96}.

\bibliographystyle{ieeetr}
\addcontentsline{toc}{section}{\refname}\bibliography{referencias}

\end{document}